\documentstyle[12pt,aaspp4]{article}
\begin{document}

\title{THE COLOR--MAGNITUDE DIAGRAM IN BAADE'S WINDOW REVISITED}

\author{Marcin Kiraga} 
\affil{Warsaw University Observatory, Al. Ujazdowskie 4, 
00--478 Warszawa, Poland}
\affil{e--mail: kiraga@sirius.astrouw.edu.pl}
\author{Bohdan Paczy\'nski and 
        Krzysztof Z. Stanek\altaffilmark{1,}\altaffilmark{2}}
\affil{Princeton University Observatory, Princeton, NJ 08544--1001, USA}
\affil{e-mail: bp@astro.princeton.edu, stanek@astro.princeton.edu}
\altaffiltext{1}{On leave from N.~Copernicus Astronomical Center, 
Bartycka 18, Warszawa 00--716, Poland}
\altaffiltext{2}{Current address: Harvard-Smithsonian Center for Astrophysics,
60 Garden Street, MS~20, Cambridge, MA 02138, USA}

\begin{abstract}

We have reanalyzed the OGLE $V$ and $I$ photometry of $\sim 500,000$
stars in Baade's Window, and we confirm the extinction map published
by Stanek (1996).  However, we find that the interpretation of the
OGLE color--magnitude diagram for the disk stars proposed by
Paczy\'nski et al. (1994) was incorrect: the dominant disk population
in Baade's Window is old, and we find no evidence for a large hole in
the inner Galactic disk.  We find evidence for a small systematic
error in the OGLE photometry for stars below the ``red clump'': the
faint stars with $V>18$ have their $(V-I)$ color indices too red by
$\sim 0.1\;mag$ .  We find tentative evidence from the OGLE and HST
photometry that the the bulge main sequence turn--off point is
brighter by $\sim 0.5 \;mag$ than it is in either 47~Tuc or NGC~6791,
indicating that the dominant population of the Galactic bulge is
considerably younger than those clusters.

\end{abstract}


\section{Introduction}

In this paper we use photometry of about 500,000 stars, measured in
the standard Johnson $V$ and Cousins $I$ over a $(40')^2$ field within
Baade's Window by the Optical Gravitational Lensing Experiment (OGLE,
cf. Udalski et al. 1992; 1993).  All the observations were made with
the 1--meter Swope telescope at the Las Campanas Observatory, operated
by the Carnegie Institution of Washington.  A catalog of 33,196 stars
in the magnitude range $14 < I < 18$ in the OGLE field BWC has
recently been published by Szyma\'nski et al. (1996).  The J2000
coordinates of the BWC field are: $\alpha = 18^h03^m24^s, ~ \delta =
-30^o02'00''$, and the field is $\sim 15'$ on a side.  The catalog
gives the $( \alpha , \delta )$ coordinates and the $I$ and $V$
magnitudes.  It will be extended in the near future to brighter and
fainter stars, as well as to much larger area.

The database of OGLE photometry (Szyma\'nski \& Udalski 1993) was used
to analyze the distribution of disk stars (Paczy\'nski et al. 1994),
to demonstrate the presence of the Galactic bar (Stanek et al. 1994;
1996), and to obtain the map of total interstellar extinction
(Wo\'zniak \& Stanek 1996; Stanek 1996).  The purpose of this paper is
to verify some of these results.

\section{Disk Stars in  Baade's Window}

\begin{figure}[t]
\plotfiddle{fig1.ps}{7.2cm}{0}{50}{50}{-160}{-75}
\caption{\small The $(V-I)-I$ color magnitude diagram for OGLE stars in
Baade's Window is shown for 10\% of the stars with $I<16.5\;mag$, with
the fraction declining to 0.7\% for the stars with $I>18.5$, to avoid
excessive crowding.  The recent OGLE catalog (Szyma\'nski et al. 1996)
lists 33,196 stars with $ 14 < I < 18$ in the field BWC, with the area
$(15')^2$, centered at $(l,b) = (1.0, -3.9)$.}
\label{fig1}
\end{figure}

The color magnitude diagram of the stars in Baade's Window shows two
distinct groups of stars, as presented in Fig.1.  The dominant
population is in the bulge (or bar), with a very prominent ``red
clump'' (or red horizontal branch) centered at $(V-I,I) \approx
(1.9,15.2)$.  The red giant branch is extended towards higher
luminosities and to the red.  The red subgiant branch is clearly
visible below the ``red clump'', and it is traceable all the way to
the main sequence turn--off point (TOP), somewhere near $(V-I,I)
\approx (1.6,18.3) $.  In addition there is a well defined and narrow
band of stars extending from $(V-I,I) \approx (0.9, 14.5)$ to
$(1.5,17.5)$, and traceable to higher luminosities.  This band has
almost the same slope as the Pleiades main sequence moved to a
distance of $\sim 2$ kiloparsecs and subject to the interstellar
extinction of $A_V \approx 1.6 $.  This coincidence led Paczy\'nski et
al. (1994) to the incorrect conclusion that there is evidence for a
large number of disk stars out to the distance $\sim 2$ kpc, and a
hole in the disk at larger distances.

The mistake caused by the slope coincidence of the ``main sequence''
was pointed out by Bertelli et al. (1995) who used their so called
HRD--GST, the Hertzsprung--Russel Diagram Galactic Software Telescope,
to identify the band with the stars located at the disk main sequence
turn--off point (TOP) at various distances along the line of sight.
The slope of the apparent sequence in the color--magnitude diagram was
determined by the variation of interstellar extinction with distance.
Initially we did not appreciate their result because we were
overwhelmed by the complexity of HRD--GST.

Rucinski (1996) demonstrated that contact binaries found in the OGLE
variable star catalogs (Udalski et al. 1994; 1995a,b; 1996) have a
uniform distribution for distances up to $\sim 8$ kpc, i.e. all the
way to the Galactic bulge, with no evidence of a major hole in the
inner Galactic disk.  The distribution of contact binaries in the
color--magnitude diagram was consistent with the Bertelli et
al. (1995) interpretation of the diagram.  Recently, Ng \& Bertelli
(1996) used the shape of the sequence made by the disk TOP stars to
study the distribution of interstellar extinction with distance.

\begin{figure}[p]
\plotfiddle{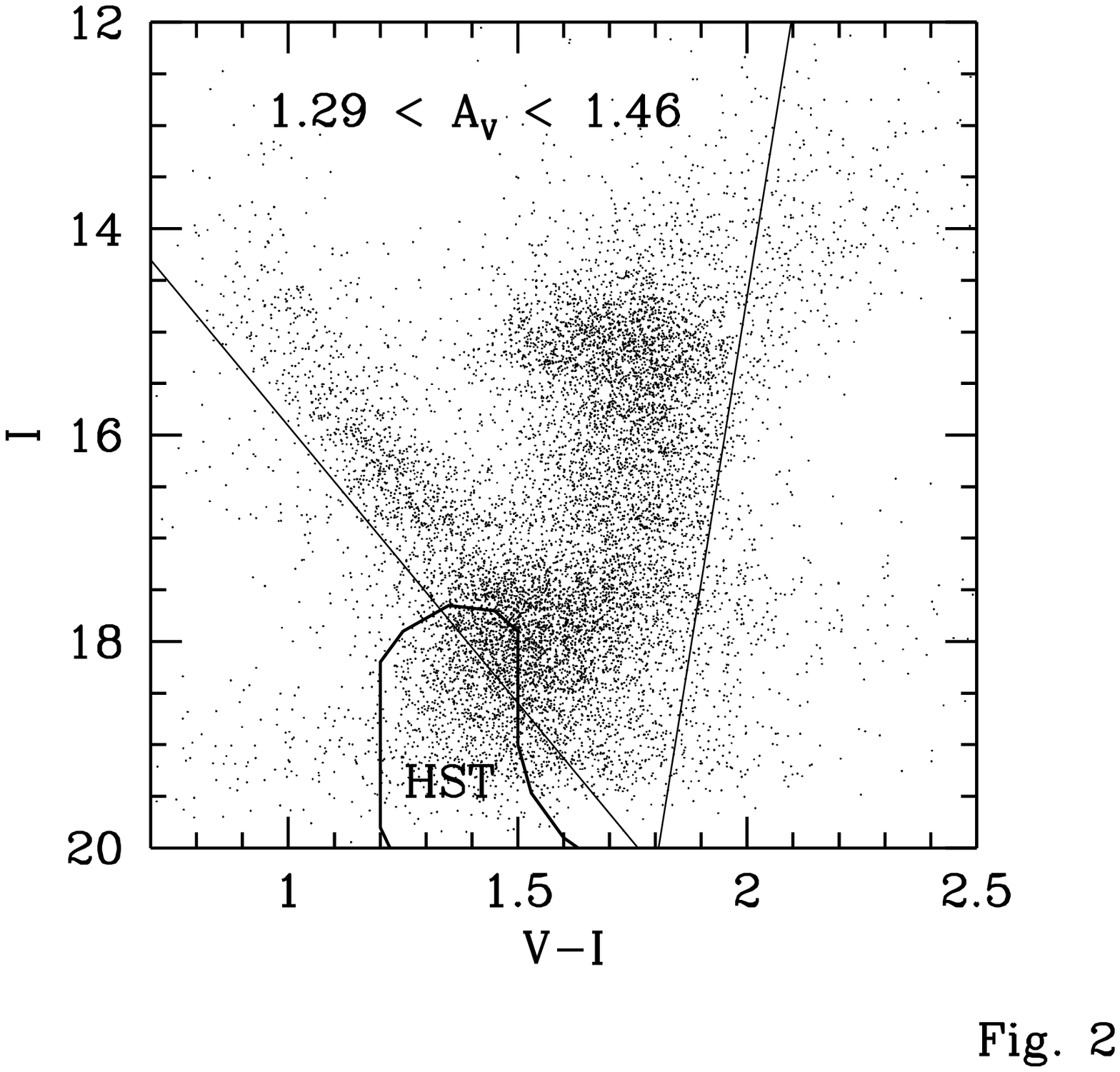}{7.2cm}{0}{50}{50}{-160}{-75}
\caption{\small The color magnitude diagram for stars in Baade's Window is
shown for the regions in which the total interstellar extinction to
the bulge is in the range: $1.29 < A_V < 1.46$.  The area at the
bottom, marked HST, indicates the location of the majority of stars
measured with the HST by Holtzman et al. (1993).  The two long solid
lines are shown for reference.  To avoid excessive crowding only a
small fraction of all stars is shown for $I > 16.9$.}
\label{fig2}
\plotfiddle{fig3.ps}{8cm}{0}{50}{50}{-160}{-85}
\caption{\small The color magnitude diagram for stars in Baade's Window is
shown for the regions in which the total interstellar extinction to
the bulge is in the range: $1.95 < A_V < 2.78$.  The area at the
bottom, marked HST, indicates the location of the majority of stars
measured with the HST by Holtzman et al. (1993).  The two long solid
lines are shown for reference.  To avoid excessive crowding only a
small fraction of all stars is shown for $I > 16.9$.}
\label{fig3}
\end{figure}

A simple demonstration of the Bertelli et al. (1995) interpretation is
provided by Figs.2 and 3, in which the color--magnitude diagrams are
shown for two subsets of stars in Baade's Window.  The two subsets
correspond to the regions in which the total interstellar extinction
towards the Galactic bulge is in the range $1.29 < A_V < 1.46$ and
$1.95 < A_V < 2.78$, respectively.  These two regions have a very
complicated structure, as determined by the extinction map provided by
Stanek (1996).  It is clear that the slope of the ``main sequence'' is
shallower in the areas of larger interstellar reddening (Fig.3).
There is also more scatter of stars in Fig.3 than in Fig.2 because of
a larger range of interstellar extinction: 0.83 mag versus 0.17 mag.
The apparent sharpness of the sequence in Fig.1 arises because most
stars in Baade's Window have a small range of extinction, with
$\langle A_V \rangle \approx 1.7\;mag$ (cf. Fig.4 of Stanek 1996).

In the following we present a simple demonstration why the
color--magnitude diagram in Baade's Window features the disk TOP stars
forming an apparently narrow sequence, while in the many directions
farther away from the Galactic plane similar stars form only a
distinct blue edge to their distribution, with $(B-V) \approx 0.4$,
and $(V-I) \approx 0.7$, but no narrow sequence is apparent in those
fields (e.g. Ka\l u\.zny \& Udalski 1992, Reid \& Majewski 1993,
Caldwell \& Schechter 1996).

A standard double--exponential disk model implies that the number
density of stars is approximately constant with distance in the
direction of Baade's Window (Paczy\'nski 1991).  At any particular
apparent magnitude the stars at the main sequence turn--off point
(TOP) can be seen out to some distance $d_{TOP} $.  Fainter and redder
main sequence (MS) stars some 2 magnitudes below the TOP can be seen
to a smaller distance $d_{MS} = 0.4 d_{TOP}$, i.e. over a volume 16
times smaller than the volume over which the TOP stars can be seen.
The number density of the MS stars is higher than the number density
of the TOP stars, but a much larger volume over which the TOP stars
are visible more than compensates for that.  The net result is that
the TOP stars dominate the color--magnitude diagram.  As the color and
absolute magnitude ranges of the TOP stars are small they form a
narrow sequence in the color magnitude diagram.  In the absence of
interstellar extinction that sequence is vertical.  The increase of
extinction with the distance makes the more distant TOP stars appear
not only fainter, but also redder, and the narrow sequence has a
distinct slope in the color--magnitude diagram, as is the case in
Baade's Window.

A sharp sequence of stars in the color--magnitude diagram is not
observed in the fields far from the Galactic plane because of rapid
decline in the number density of disk stars with the distance, which
reduces the advantage of a larger volume over which the TOP stars are
detectable.  While the TOP stars are still responsible for the
sharpness of the blue edge in those fields, the more numerous and
redder MS stars are responsible for a broad and diffuse distribution
to the red.

\begin{figure}[t]
\plotfiddle{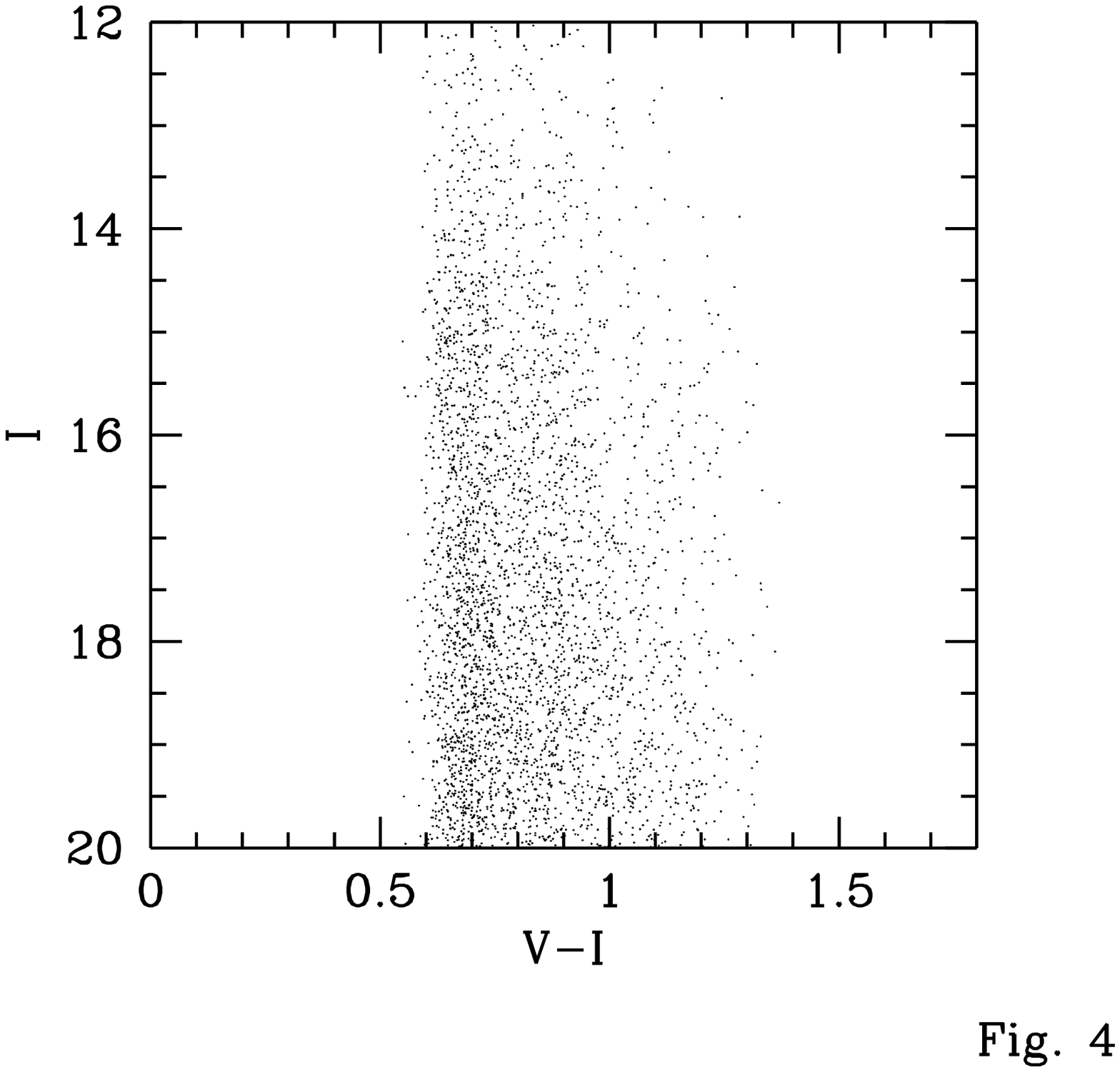}{7.2cm}{0}{50}{50}{-160}{-75}
\caption{\small The color magnitude diagram for the old disk population made
of 47~Tuc stars, with the number density falling off with distance
(cf. Eq.1).  No interstellar extinction has been included in this
model.  Notice the sharp blue edge of the distribution, similar to
that observed in many direction in the sky, far from the Galactic
plane (cf. Ka\l u\.zny \& Udalski 1992, Reid \& Majewski 1993,
Caldwell \& Schechter 1996).}
\label{fig4}
\end{figure}

\begin{figure}[t]
\plotfiddle{fig5.ps}{7.2cm}{0}{50}{50}{-160}{-75}
\caption{\small The color magnitude diagram for the old disk population made
of 47~Tuc stars, and the number density constant out to $5.6\;kpc$.
Beyond $5.6\;kpc$ the number density is given by the E2 model for the
Galactic bar, proposed by Dwek et al. (1995) and modified by Stanek
et al. (1996). The fraction of model stars shown is reduced smoothly
between $I=16.5\;mag$ and $I=19.3\;mag$ by a factor of 50, to avoid
excessive crowding in the lower part of the diagram.  The variation of
the interstellar extinction with distance is given with the eq. (2).
Notice the narrow and inclined sequence of stars between $(V-I,I) =
(0.9, 14) $ and $(1.5,18)$, which is similar to that found by the
OGLE in Baade's window (cf. Fig.1).  This sequence is made of the disk
TOP stars, and its shape is governed by the variation of interstellar
extinction with distance (cf. Bertelli et al. 1995, Ng et al. 1995, Ng
\& Bertelli 1996).}
\label{fig5}
\end{figure}

A simple visualization of the effect is presented in Figs. 4 and 5.
Following Reid \& Majewski (1993) we took the stars in the globular
cluster 47~Tuc (Ka\l u\.zny et al. 1996) as an approximate
representation for the thick disk color--luminosity function.  We
considered two forms of number density variation with distance.
First, the number density was assumed to vary as
\begin{equation}
n = n_0 \left( 1 +  { d \over 500 ~ pc } \right) ^{-3} , 
\end{equation}
where $n_0$ is the local number density near us.  We also assumed
there was no interstellar extinction.  As a result we obtained a
color--magnitude diagram presented in Fig.4.  The sharp blue edge is
clearly apparent, but there is no striking red edge.  This simulation
provides a similar visual impression as the observed color--magnitude
diagrams of Ka\l u\.zny \& Udalski (1992), Reid \& Majewski (1993) and
Caldwell \& Schechter (1996).

In order to simulate the conditions in Baade's Window we adopted the
same 47~Tuc color -- luminosity function, but we assumed that the
number density of disk stars is constant between us and 5.6 kpc, and
that it follows the number density of stars given by the E2 model for
the Galactic bar, proposed by Dwek et al. (1995) and modified by
Stanek et al. (1996), at distances larger than 5.6 kpc.  The
interstellar extinction as a function of distance $d$ was adopted to
be:
\begin{equation} 
A_V = 2 ~ mag ~ \left[ 1 - e^{-(d/1.7 ~ kpc )^2} \right] ,
\end{equation}
i.e. somewhat different than recommended by Arp (1965).  The
corresponding color--magnitude diagram is shown in Fig.5.  The
sequence of stars from $(V-I, I) = (0.9, 14.0)$ to $(1.5, 18.0)$ has a
resemblance to the observed sequence (cf. Fig.1), and to the model
results presented by Bertelli et al. (1995).  This sequence is made of
the disk TOP stars.  Note that the sequence becomes vertical for $ I >
18.2$, as there is almost no additional extinction (and reddening)
beyond $ \sim 3 \;kpc$ in the model.  The sequence remains narrow
along its full length in Fig.5, for $ 14 < I < 20$, with the lowest
part dominated by the Galactic bulge/bar stars.

A comparison between Figures 1 and 5 makes it clear that our single
population model cannot provide a quantitative explanation for the
many details of the observed color--magnitude diagram.  However, our
goal was to provide a simple explanation of the origin of the sharp
sequence of disk stars in Baade's Window, and the reason for its
absence in the other directions.  We have shown that the sharp
sequence appears because the number density of disk stars remains
roughly constant with distance in Baade's Window, while it declines
rapidly in the other direction.

The many details of the observed color--magnitude diagram as presented
in Fig.1 depend on the mix of many stellar populations, each with its
own range of chemical compositions, its range of ages, its geometrical
distribution, its own mass function, and therefore its specific
luminosity function.  In addition, there is a complicated variation of
the interstellar extinction with distance.  According to Bertelli et
al. (1995) and Ng et al. (1996), their multi--component HRD--GST can
explain most of the observed details, but this is at present beyond
our ability to asses.

The OGLE data cover only two photometric bands and it should not be
over--interpreted.  It is necessary to take into account the
incompleteness as described by Szyma\'nski et al. (1996) before making
quantitative inferences.  Also, there seems to be a systematic error
in the colors of stars fainter than V $ \sim 18$, as discussed in the
following section of this paper.  That error will be corrected with a
new data set to be obtained shortly with a new instrument.

\section{Galactic Bulge/Bar}

The line of sight through Baade's Window reaches the distance of 8 kpc
at $ \sim 500$ pc below the Galactic plane.  It is most likely that
there is no significant extinction so far out.  According to Arp
(1965) the extinction ends at a distance $ \sim 2 $ kpc away from us,
while according to Ng \& Bertelli (1996) it may extend out to $ \sim 4
$ kpc.  In any case it is safe to assume that the inner disk, as seen
through Baade's Window, and the stars in the Galactic bulge/bar
farther out, are subject to the same total extinction map (Stanek
1996).  Therefore, in order to study the inner parts of the galaxy we
can make a correction following Stanek's prescription.  We applied the
extinction corrections to all stars, which means we have
over--corrected the disk stars which are closer to us than $ \sim 4 $
kpc.  The corresponding color--magnitude diagram is shown in Fig.6,
where only 10\% of bright stars and 0.7\% of the faint stars were
plotted, to avoid excessive crowding.

\begin{figure}[t]
\plotfiddle{fig6.ps}{7.2cm}{0}{50}{50}{-160}{-75}
\caption{\small The $(V-I)_0 - I_0$ color magnitude diagram for OGLE stars
in Baade's Window is shown for 10\% of the stars with $ I < 16.5 $
mag, and the fraction declining to 0.7\% for the stars with $ I > 18.5
$, to avoid excessive crowding.  The reddening correction was applied
following Stanek (1996).  It is correct for the bulge stars, but it is
too large for the bright disk stars, i.e. the disk sequence has been
made too blue.  The diagrams for 47~Tuc (Ka\l u\.zny et al. 1996), and
NGC~6791 (Ka\l u\.zny \& Udalski, 1992), corrected for interstellar
reddening, and shifted to the distance of 8 kpc, are shown
schematically with thick solid lines, The sequence of stars measured
with the HST (Holtzman et al. 1993), and corrected for interstellar
reddening following Stanek (1996), is shown with dashed lines.  The
average color of the faint OGLE stars is shown with open circles.}
\label{fig6}
\end{figure}

The outlines of stellar sequences observed in the globular cluster
47~Tuc, and in the very old, and metal rich open cluster NGC~6791, are
shown with solid lines in Fig.6.  The cluster sequences were corrected
for reddening.  We adopted $ E_{V-I} = 0.06$ for 47~Tuc following Ka\l
u\.zny et al. (1996), and $ E_{V-I} = 0.22$ for NGC~6791 following
Ka\l u\.zny \& Udalski (1992).  Both sequences were shifted in
magnitude as if moved to the distance of 8 kpc, adopting $ (m-M)_V =
13.4 $ and $ (m-M)_V = 13.52 $ for 47~Tuc and NGC~6791, respectively
(i.e. $ (m-M) = 13.25 $ and $ (m-M) = 12.97$, respectively).  The area
surrounded with a dashed line at $I_0 > 16.8$ and $0.63 < (V-I)_0 <
1.0$ is a schematic representation of the star sequence observed with
the HST (Holtzman et al. 1993) and corrected for extinction following
Stanek (1996): $E_{V-I} = 0.56$, which is almost identical to $E_{B-V}
= 0.46$ adopted by Holtzman et al. (1993).  A series of open circles
at $I_0 > 16.6$ and $(V-I)_0 \approx 0.95$ corresponds to average
colors of faint OGLE stars as a function of magnitude.

The adopted relative distances to 47~Tuc, NGC~6791, and the Galactic
bulge may be somewhat incorrect, and the same is true about the
reddening.  The ages and metallicities of all three groups of stars:
47~Tuc, NGC~6791, and the Galactic bulge, are subject to some
uncertainty too.  According to stellar model calculations
(cf. Bertelli et al. 1994) the luminosity of red clump (horizontal
branch) stars varies with age and metallicity.  Therefore, we make no
attempt to refine the relative positions of the three red clumps in
Fig.6, but instead we consider the implications of some features of
the star sequences which are distance independent.

One of the reddening and distance independent properties is the slope
of the red subgiant sequence for $ 14.5 < I_0 < 16.5 $.  The observed
OGLE sequence is almost vertical, while the subgiant branches of
47~Tuc and NGC~6791 have a distinct slope, with the brighter stars
being redder.  This suggests that there may be a magnitude dependent
color error in the OGLE photometry.

The colors of TOP stars in both clusters are similar, and they are not
very different from the colors of bulge TOP stars obtained with the
HST (Holtzman et al. 1993), but they are all much bluer than the
average colors of OGLE stars at the same magnitude, $ I_0 \approx 18$.
This indicates that there is some systematic error in the OGLE
photometry, as the faint OGLE stars appear too red.

In fact there was some evidence of a non--linearity of the CCD camera
system used by the OGLE (Udalski et al. 1993).  The case seems to be
strengthened with a comparison between the OGLE photometry of the TP8
field (at $ l=-0.1, ~ b=-8.0$) with the recent photometry done on the
2.5--meter Du Pont telescope by Ka\l u\.zny \& Thompson (1996).  It
indicates a systematic color shift of $ \sim 0.1\;mag$ , in the sense
that faint OGLE stars are too red.

Note, that for the faintest OGLE stars the systematic effects of
stellar image blending is significant.  Most stars below the bulge TOP
are the bulge MS stars, which are redder.  Therefore, the measured
colors of the unresolved blended images may be systematically too red,
adding to the systematic effect caused by a non--linear response of
the CCD detector and the camera electronic.

A much better photometry with the new OGLE--2 telescope is expected to
be available soon.  Therefore, we think it is not useful to
investigate the systematic error in the old OGLE photometry.  It had
no effect on the microlensing searches or the search for stellar
variability.  However, the presence of a systematic error limits the
use of the old OGLE color--magnitude diagrams for a quantitative
analysis of stars with $ V > 18$, or $ I > 16 $.

\section{The Bulge/Bar Age}

There is an intriguing difference between the OGLE color--magnitude
diagram for the Galactic bulge/bar stars, and the diagrams for 47~Tuc
and NGC~6791 at $ I_0 \approx $ 16.7 -- 17.7, $(V-I)_0 \approx $ 0.8
-- 1.0 (cf. Fig.6).  There is a short, almost horizontal segment
joining the TOP region with the red subgiant branch.  In the two
clusters the magnitude difference between this segment and the red
clump at $(V-I)_0 = 1.05$ is $ \sim 3.2\;mag$ , while the OGLE data
indicate the difference to be only $ \sim 2.6\;mag$ .  This
difference, just like the better known (but more difficult to measure)
magnitude difference between the horizontal branch stars and the TOP
stars, is a good age indicator (cf. Bertelli et al. 1994).  Taken at
the face value it implies that the dominant bulge/bar population is
substantially younger than the clusters 47~Tuc and NGC~6791.  However,
considering the problems with the systematic color errors and blending
for faint OGLE stars, and only marginal evidence for the presence of
the ``horizontal segment'' in the OGLE data as presented in Fig.6, the
evidence for the age of the bulge is marginal at best.


The same ``horizontal segment'' in the directly observed $(V-I)$--$I$
diagram is marginally apparent in Fig.1 at $ I \approx 17.8$, $(V-I)
\approx 1.7 $.  The HST data presented in Fig.2 of Holtzman et
al. (1993) has too few stars at $ I < 17.8$ to make any claim about
the presence or absence of the segment at $ I \approx 17.8 $.
However, with the center of the red clump at $ I \approx 15.3$, the
``horizontal segment'' would be expected to be at $ I \approx 15.2 +
3.2 = 18.4$ if the bulge was like 47~Tuc or NGC~6791, but there is no
such thing apparent in the HST data at $ I \approx 18.4 $.
Unfortunately, the number of stars in the HST color--magnitude diagram
is too small to be confident.

Therefore, rather inconclusive HST data seem to point the same way as
the inconclusive OGLE data: the shape of the bulge/bar
color--magnitude diagram between the TOP and the red subgiant branch
seems to imply a substantially lower age for the dominant bulge
population than the age of old globular or open clusters.  The
unusually rich red clump (red horizontal branch) of the bulge/bar
provides yet another indication of a relatively young population
(Paczy\'nski et al. 1994).

The issue of the bulge age is vigorously discussed in the literature,
and it is not settled yet.  It could be resolved with the HST
color--magnitude diagram extending to bright enough stars, and over
large enough area, to cover the star sequence joining the TOP with the
red subgiant branch.  Even more definite resolution of the age problem
could be provided by the determination of masses of detached eclipsing
binaries in that part of the diagram (Paczy\'nski 1996).  Such
binaries can be found in the OGLE catalogs (Udalski et al. 1994,
1995a,b, 1996).  The main practical problem is the faintness of those
stars; the radial velocity curves may have to wait until the 6.5 meter
and 8 meter telescopes become operational.

\section{Interstellar Extinction Map}

Stanek (1996) has used the method developed by Wo\'zniak \& Stanek
(1996) to construct the extinction map for Baade's Window.  The method
took advantage of red clump stars as tracers.  In this section we
verify the map using two other groups of stars: the red subgiants and
the far disk and bulge TOP stars.  The regions of the color--magnitude
diagram covered by the three groups is shown in Fig.7 with the three
parallelograms, which have the slope of the upper and lower boundaries
fixed by the relation that holds for interstellar extinction: $ E(I) =
1.5 \times E(V-I) $.

\begin{figure}[t]
\plotfiddle{fig7.ps}{7.2cm}{0}{50}{50}{-160}{-75}
\caption{\small The $(V-I)_0 - I_0$ color magnitude diagram for OGLE stars
in Baade's Window, together with the three boxes indicating the
location of the red clump stars, red subgiant stars, and TOP stars,
which were used for the three independent estimates of interstellar
reddening.}
\label{fig7}
\end{figure}

All stars in the OGLE database for Baade's Window were ordered
according to the value of their extinction using Stanek's (1996)
software, and divided into 20 groups with equal number of stars in
each group.  Next, the stars in the three regions as shown in Fig.7
were selected for each of the 20 groups.  Finally, the stellar
$(V-I)_0$ and $ I_0$ values were shifted for each group along the
extinction lines so as to achieve the best match between all the
groups.  This way residual corrections were found to the average
values of interstellar extinction for each reddening group, and for
each type of star: the red clump stars, the red subgiants, and the TOP
stars.  If these were perfect methods to measuring interstellar
extinction then all the corrections should be close to zero, and there
should be no systematic trend between them.

The comparison between the extinction values obtained for each of the
20 reddening groups for all three types of stars is shown in Fig.8,
together with the dashed line which indicates the perfect relation.
The overall agreement is reasonable, but a small but systematic trend
is disturbing.  It is most disturbing for the red clump stars which
are shown with the small filled circles.  The range of reddening
values as determined in this paper (KPS) is larger than the range
obtained from the Stanek's map.

\begin{figure}[p]
\plotfiddle{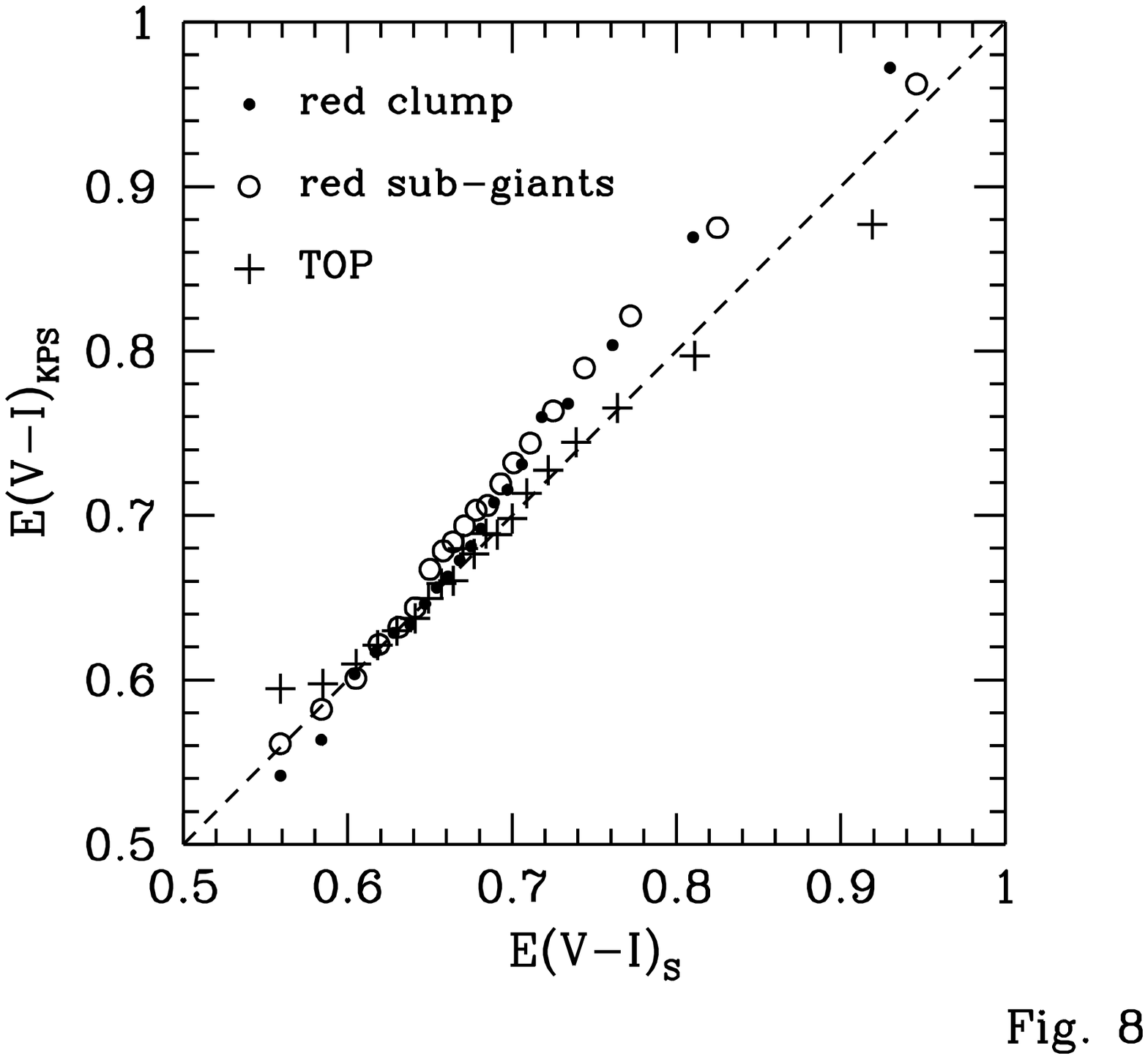}{7.2cm}{0}{50}{50}{-160}{-75}
\caption{\small The relation between the interstellar reddening as determined
in this paper $ [E(V-I)_{KPS}]$ and as given by Stanek (1996), $
[E(V-I)_S]$, is shown for three groups of stars.  The location of the
three groups in the color--magnitude diagram is shown with three
boxes in Fig.7.}
\label{fig8}
\plotfiddle{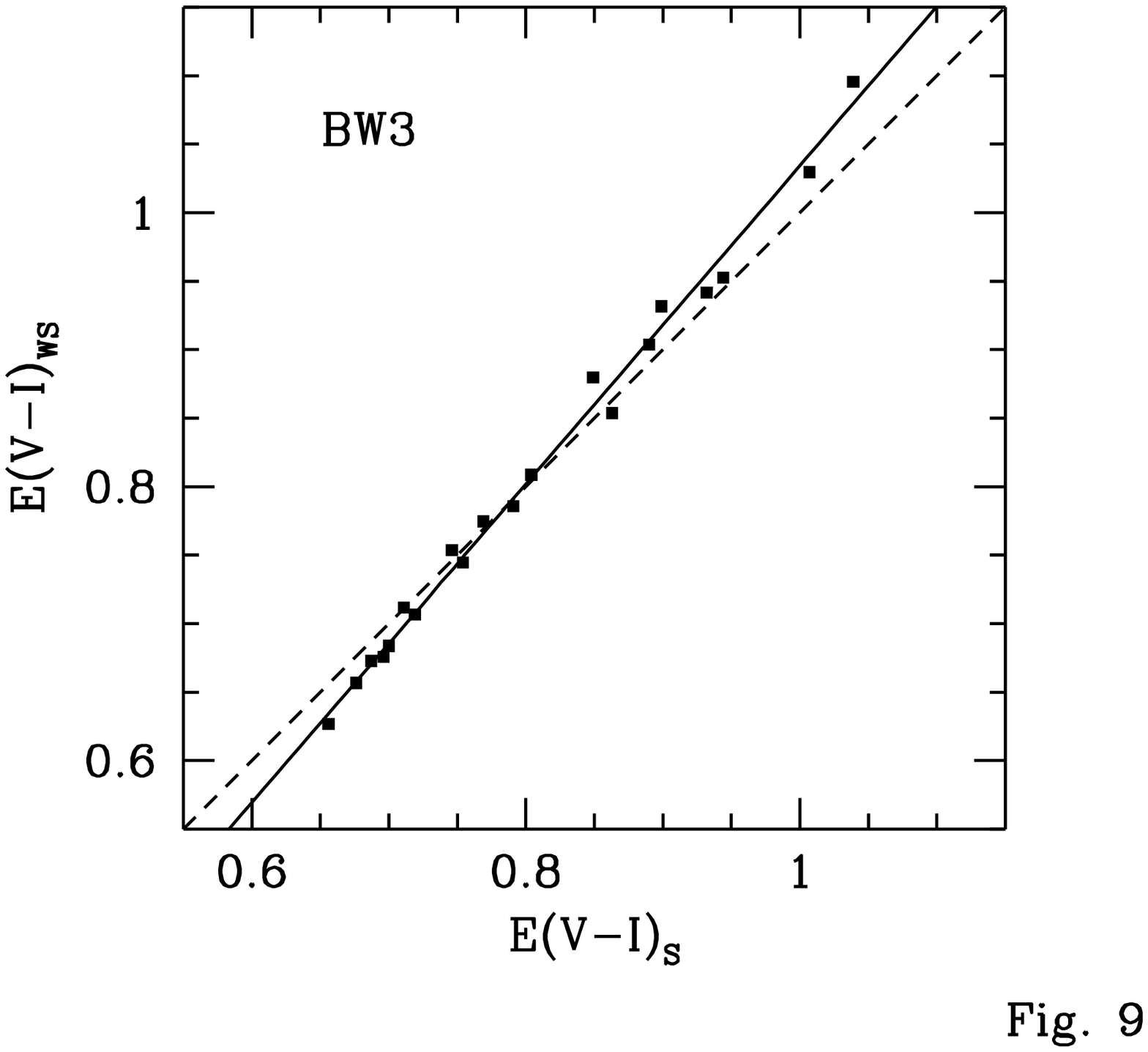}{8cm}{0}{50}{50}{-160}{-85}
\caption{\small The relation between interstellar reddening determined by
Wo\'zniak \& Stanek (1996),  $[E(V-I)_{KPS}]$, and given by Stanek
(1996), $[E(V-I)_S]$, is shown stars in the OGLE fields BW3.  The
small dark squares correspond to average values for 20 groups of stars
with similar values of $ [E(V-I)_S]$, with the solid line drawn
following the least square solution.  The slope of the solid line is
not $45^o$ because of interpolation used by Stanek (1996).}
\label{fig9}
\end{figure}

It turns out that a significant fraction of the systematic trend
apparent in Fig.8 is caused by the interpolation, as described by
Stanek (1996).  This is demonstrated in Fig.9, where the estimate of
the reddening for all stars in the OGLE database in the field BW3 is
shown for 20 groups ordered according to the value of their extinction
(the field BW3 was selected because it has the largest range of
extinction).  The value of $ E(V-I)_S$ was obtained with the Stanek's
extinction map, and it involved the interpolation between the original
subframes for which the values of $ E(V-I)_{WS}$ were estimated by
Wo\'zniak \& Stanek (1996).  The data on which the extinction map was
based had the resolution limited by these subframes, which were $ 28''
$ on a side.  The difference between the perfect system (dashed line
in Fig.9) and the real world (solid line) is in the same sense and
about the same magnitude as the systematic trend present in Fig.8.
However, some part of the trend may be caused by the systematic errors
in the OGLE photometry, as discussed in section 3.

The fact that the values of interstellar reddening estimated with
three different types of Galactic bulge stars are all in a fair
agreement, as shown in Fig.8, demonstrates that Stanek's (1996)
reddening map can be used with some confidence.  On the other hand the
differences apparent in Fig.8 indicate how large the differential
errors of the map are.  Of course, there is a separate issue of the
reddening zero point, as discussed by Stanek (1996).

\section{Discussion and Conclusions}

This paper addresses two errors we have found in the previous analysis
of the OGLE data, and in the data itself: the incorrect interpretation
of the disk star sequence in the color--magnitude diagram by
Paczy\'nski et al. (1994), and a systematic photometric error of $
\sim 0.1 \;mag$ for the faint stars.  The correct interpretation of
the disk sequence as made of stars at the main sequence turn--off
points at various distances, and suffering interstellar reddening
increasing with the distance, was first proposed by Bertelli et
al. (1995).  In this paper we explained the sharpness of the sequence
as caused by the fact that the number density of stars seen through
Baade's Window remains roughly constant as a function of distance.
This is approximately the same population which dominates the
color--magnitude diagrams far from the Galactic plane, and creates the
characteristic blue edge at $(B-V) \approx 0.4$, and $(V-I) \approx
0.7$ (e.g. Ka\l u\.zny \& Udalski 1992, Reid \& Majewski 1993,
Caldwell \& Schechter 1996).

The photometric error is most likely caused by a non--linearity of the
detector system, as mentioned by Udalski et al. (1993), and confirmed
by Ka\l u\.zny \& Thompson (1996), with some additional contribution
due to crowding and blending of the faint stars with the bulge main
sequence stars which are below the detection threshold, and which are
systematically redder than the stars just above the detection
threshold.  We expect that photometry from the new OGLE--2 system will
not be subject to such systematic errors.

We verified the extinction map based on the red clump stars, obtained
by Stanek (1996), using the bulge red subgiant stars and the bulge TOP
stars.  We found reasonable agreement between these groups of stars,
but we have also found a small systematic error caused by the
interpolation process which is used by the map software.  At this time
Fig.8 can be used to estimate the map accuracy.

We found some evidence in the OGLE data, and in the HST data as
presented by Holtzman et al. (1993), that the dominant bulge
population is considerably younger than either moderately metal weak
globular cluster 47~Tuc, or the the metal rich open cluster NGC~6791.
This relatively young age as indicated by the location of stars
between the TOP and the red subgiant branch, barely detectable in
Fig.6.  A definite confirmation of the effect can be obtained with new
HST data which must cover a larger area and extend to brighter stars
than those analyzed by Holtzman et al.  (1993), so that the location
of the star sequence connecting the bulge TOP with its red subgiant
branch can be clearly established.

\acknowledgments{We want to thank Wes Colley for this careful reading
of this manuscript. It is also a great pleasure to acknowledge that
all figures in this paper were obtained with the SM graphics software
developed by R. H. Lupton.  This project was supported with the NSF
grants AST--9216494 and AST--9528096.  The work of MK was also
supported with the KBN grant 2 P304 004 007.}


\end{document}